\documentclass[12pt,preprint]{aastex}

\slugcomment{To appear in the Astrophysical Journal}

\shorttitle{Energetics of a Flaring Active Region}
\shortauthors{Wheatland}

\begin{document}

\title{The Energetics of a Flaring Solar Active Region, and Observed
Flare Statistics}

\author{M.S. Wheatland} \affil{School of Physics, University of Sydney,
NSW 2006, Australia}

\begin{abstract} 
A stochastic model for the energy of a flaring solar
active region is presented, generalising and extending the approach
of \citet{whe98}. The probability distribution for the free energy of an
active region is described by the solution to a master equation involving 
deterministic energy input and random jump
transitions downwards in energy (solar flares). It is shown how two
observable distributions, the flare frequency-energy distribution and
the flare waiting-time distribution, may be derived from the
steady-state solution to the master equation, for given choices for the
energy input and for the rates of flare transitions. An efficient method
of numerical solution of the steady-state master equation is presented.
Solutions appropriate for flaring, involving a constant rate of energy
input and power-law distributed jump transition rates, are
numerically investigated. The flare-like solutions exhibit power-law
flare frequency-energy distributions below a high energy rollover,
set by the largest energy the active region is likely to have. 
The solutions also exhibit approximately exponential (i.e.\ Poisson)
waiting-time distributions, despite the rate of flaring depending on the
free energy of the system. 
\end{abstract}

\keywords{Sun: flares --- Sun: corona --- Sun: activity --- Methods:
statistical }

\section{Introduction}\label{intro}

Solar active regions are the sites of occurrence of most solar flares.
Large active regions may persist for several solar rotations, and
produce dozens of significant flares during a transit of the disk
\citep{ric51}. The largest flares are believed to involve the
release of more than $10^{27}\,{\rm J}$ of stored magnetic energy, the
energy appearing in accelerated particles, heating, bulk motion of
material, and radiation \citep{hud91}. The process underlying flares is 
accepted to be magnetic reconnection, although the details of the process 
remain the subject of study \citep{pri02}.

The dynamical energy balance of solar active regions presents challenges 
to our understanding. The energy liberated in flares is thought to be 
excess
or `free' magnetic energy associated with electric current systems in the 
solar atmosphere, but the origin of the currents, and hence the source of 
the energy, is not well understood \citep{mcc87,tan88,lek96,whe00a,sch05}. 
Two popular
pictures for energy supply are first that existing coronal magnetic
structures are twisted and sheared by photospheric motions, producing
currents, and second that new current-carrying magnetic flux emerges 
through the photosphere.
These mechanisms suggest that energy supply to an active region may be
described as a continuous process, driven by slow photospheric and
sub-photospheric motions. In contrast, flare energy release is rapid and
unexpected. An important point is that the size of the downwards jump in
energy associated with a flare may be very large, by comparison with the
amount of stored active region energy.

The understanding of active region energy balance is hampered by the
inability to calculate coronal magnetic energy. In principle the 
magneto-hydrodynamic (MHD) virial theorem permits the calculation of 
magnetic energy from vector field values inferred in the chromosphere 
subject to the assumption that the field is everywhere force free 
(e.g.\ Metcalf et al.~2005). However few chromospheric vector field 
determinations are made, and the reliability of the method is unknown. 
Methods for modeling coronal magnetic fields 
from photospheric or chromospheric boundary conditions are being 
developed \citep{met08,sch08}, and these may permit estimation of coronal 
magnetic energy. There has also been recent progress in methods for 
estimating the rate of supply of energy to an active region directly from 
observations (e.g.\ Welsch et al.\ 2007). For example, in ideal MHD the 
Poynting flux is ${\bf S}=(-{\bf u}\times {\bf B})\times {\bf B}/\mu_0$, 
where ${\bf u}$ is the fluid velocity and ${\bf B}$ is the magnetic 
field, and in principle this quantity may be estimated at the photosphere 
from observations.

Although we lack detailed quantitative information about the rate of
energy supply to, and the energy stored in, active regions, we do have
detailed information about solar flare occurrence. The energy released
in flares may be estimated (albeit subject to some error), and the rate of
occurrence of flares is observed. Two related statistical properties of
flares -- the frequency-energy distribution, and the distribution of
times between events -- have been studied in some detail, as summarized
below.

Studies of the frequency-energy distribution show that it is a power law
over many decades in energy \citep{hud91,cro93,asc98}. Specifically the
distribution may be written 
\begin{equation} 
{\cal N}(E)=AE^{-\gamma},
\end{equation} 
where ${\cal N}(E)$ is the number of flares per unit time
and per unit energy $E$, the factor $A$ is a (time-dependent) measure of
the total flaring rate, and $\gamma\approx 1.5$. Typically this
distribution is determined for flares from all active regions on the Sun
over some period of time, but it also appears to apply to individual
active regions \citep{whe00b}, which suggests that the power law is
intrinsic to the flare mechanism. A popular model explaining the power
law is the avalanche model \citep{lu91,cha01}, in which the magnetic
field in the corona is assumed to be in a self-organized critical state,
and subject to avalanches of small-scale reconnection events.

There has been considerable interest in the flare waiting-time
distribution, and more generally in waiting-time distributions for what 
they reveal about underlying physics in a variety of systems 
\citep{san02}.
Determinations of flare waiting-time distributions have given varied 
results \citep{pea93,bie94,whestu98,bof99,moo01,whe01,moo02}. The results 
suggest that the observed distribution depends on the particular active 
region and on time, and that it is also influenced by event 
definition and selection procedures \citep{whe01,buc05,pac05,bai06}. 
For some active regions, the distribution appears to be consistent 
with a simple Poisson process, i.e.\ independent events occurring at a 
constant mean rate \citep{moo01}. The corresponding waiting-time 
distribution is exponential. Other active regions show time-variation 
in the flaring
process, and flare occurrence may be approximated by a piecewise
constant, or more generally time-varying Poisson process \citep{whe01}.
On longer time scales a power-law tail is observed for events from the
whole Sun \citep{bof99}. This may be accounted for in terms of a
time-dependent Poisson model \citep{whe02}, although some authors have
argued that the power law has fundamental significance \citep{lep01}.
We note that the waiting-time distribution has often been considered in 
isolation from other flare statistics, in particular the frequency-energy
distribution.

To understand the statistics of flare occurrence, it is desirable to
have a general theory for the energetics of an active region, relating the
free energy of the active region to the observed frequency-energy and
waiting-time distributions. \citet{ros78} presented the first model of
this kind, which was analogous to Fermi acceleration. In this model
active regions experience exponential growth in free energy between 
flares which occur as a Poisson process in time, depleting all accumulated
energy. This gives a power-law flare frequency-energy distribution for
large energies, and presupposes Poisson occurrence. Litvinenko (1994;
1996) generalized the model to incorporate different rates of supply of
energy to the system, but the Litvinenko models retained the 
feature that each flare releases all of the free energy. This aspect of
the \citet{ros78} model was criticized by \citet{lu95}. \citet{whe98}
introduced a model permitting arbitrary changes in free energy at each
flare. The model assumes the free energy $E$ of an active region 
increases secularly between jump transitions downward in energy 
(flares), which occur at a rate $\alpha (E,E^{\prime})$ for jumps from 
$E$ to $E^{\prime}$, per unit energy.
A master equation describes the steady-state energy balance, and the 
solution of this equation is the probability distribution $P(E)$ 
for the free energy. The flare frequency-energy distribution is given 
by the convolution
\begin{equation}\label{eq:ffe} 
{\cal N}(E)=\int_{E}^{\infty}P(E^{\prime})\alpha 
(E^{\prime},E^{\prime}-E)dE^{\prime}. 
\end{equation} 
\citet{whe98} investigated solutions to the
master equation for constant energy input and for a choice of
transition rates $\alpha (E,E^{\prime})\sim (E-E^{\prime})^{-\gamma}$, 
which leads 
to power-law behavior ${\cal N}(E)\sim E^{-\gamma}$ (below a high 
energy rollover set by the highest energy an active region is likely 
to have). \citet{whe98} did not determine a waiting-time distribution 
for the model. It was argued that the derived power-law solutions are 
consistent with an avalanche-type model, and avalanche models have 
simple Poisson waiting-time statistics \citep{whestu98,san02}. 
However, this presents a puzzle: the total rate of flaring in the 
\citet{whe98} model is given by
\begin{equation}\label{eq:lambda} 
\lambda (E)=\int_0^{E}\alpha
(E,E^{\prime})dE^{\prime},
\end{equation} 
 which depends on the energy of the system. Hence it is
expected that the occurrence of flares is not strictly Poisson, since
the occurrence of a flare changes the energy of the system, and hence
the instantaneous total rate of flaring. Non-Poisson waiting-time 
statistics might then be expected.

Recently \citet{dal07} demonstrated how to determine steady-state 
waiting-time distributions for continuous time processes with arbitrary 
jump transitions. The \citet{dal07} theory is quite general, 
applying to any system described by a single time-dependent stochastic 
variable $x(t)$ following a deterministic trajectory interrupted by 
positive or negative jumps of random timing and size. The probability 
distribution $P(x,t)$ for $x(t)$ is described by a master equation, and
the waiting-time distribution for jumps in the steady state may be 
obtained from the solution to the steady-state master equation.
\citet{dal07} demonstrated the theory in application to simple 
models for human attention, for voltage across a nerve membrane, and 
for soil moisture content associated with rainfall events. In each case 
the models were analytic, involving simple solutions to the master equation. 
In this paper the \citet{dal07} theory is applied to the \citet{whe98} 
model for active region free energy, and waiting-time distributions are 
derived for solutions of relevance to flares. The application of the theory 
is relatively straightforward: the Wheatland \& Glukhov (1998) model is 
in the class of models considered by \citet{dal07}, the time-dependent
stochastic variable being the active region energy $E(t)$. Minor
modifications to the theory are required because $E$ is a positive 
definite quantity, and because the jump transitions in $E$ are always 
negative. A specific difficulty in applying the theory is that for the 
power-law form for rate transitions relevant for flares, the master 
equation is not amenable to analytic solution. 
In this paper an efficient 
numerical method of solution of the steady state master equation is 
presented and applied. This paper also considers a more general form 
for flare transition rates than considered in \citet{whe98}. The results 
resolve the puzzle outlined above concerning whether the model produces an
exponential waiting-time distribution. 

The sections of the paper are as follows.
Section~\ref{model} presents the model, starting with the time-dependent 
master equation (\S\,\ref{master}).  Section~\ref{wtd} shows how
steady-state waiting-time distribution may be obtained, 
\S\,\ref{analytic_solns} presents simple analytic solutions illustrating 
the theory, and \S\,\ref{moments} considers the information provided by 
moments of the master equation.
Section~\ref{flare_solns} presents flare-like solutions to the master
equation, starting with a justification of appropriate choices for the
rates of transitions and for the energy supply rate 
(\S\,\ref{flare_choices}).
Section~\ref{numerical} describes the numerical method, and
\S\,\ref{results} presents the results. Section~\ref{discussion}
discusses the results, and their significance for understanding solar 
flares.

\section{Model}\label{model}

\subsection{Master equation} \label{master}

An active region is modeled as a system with free energy $E(t)$ which
evolves in time due to secular energy input and jumps downward in energy
at random times and of random sizes. The system is described by the
time-dependent master equation \citep{vankam92,gar04} for the
probability distribution $P(E,t)$ of the free energy:
\begin{eqnarray}\label{eq:master} 
\frac{\partial P(E,t)}{\partial t} =
&-&\frac{\partial }{\partial E}\left[\beta (E,t) P(E,t)\right]
-\lambda (E)P(E,t)
\nonumber \\ &+&\int_E^{\infty}P(E^{\prime},t)\alpha
(E^{\prime},E)dE^{\prime}, 
\end{eqnarray} 
where $\beta (E,t)$
describes the energy input rate at time $t$, $\alpha (E,E^{\prime})$
describes the rate of flare jumps from $E$ to $E^{\prime}$, and
$\lambda (E)$ is the total rate of flaring, given by
equation~(\ref{eq:lambda}). The terms on the right hand side of
equation~(\ref{eq:master}) describe the system gradually increasing in
energy due to energy input, falling to a lower energy due to a flare,
and falling from a higher energy due to a flare, respectively. This is
the time-dependent version of the master equation given in
\citet{whe98}.

Following \citet{dal07}, we note that the system is also described by 
the stochastic differential equation 
\begin{equation}\label{eq:stoch_de}
\frac{dE}{dt}=\beta (E,t)-\Lambda (E,t) 
\end{equation} 
where $\Lambda (E,t)=\sum_{i=1}^{N(t)}\Delta E_i\delta (t-t_i)$ 
describes accumulated
losses in energy due to flaring, $\delta (x)$ is the delta function, and
where the times $t_i$ are given by a state-dependent Poisson process
with occurrence rate $\lambda [E(t)]$. The jump amplitudes $\Delta E$
are distributed according to the (state-dependent) distribution 
$h (\Delta E,E)$, defined by 
\begin{equation}\label{eq:h} 
\alpha (E,E-\Delta E )=\lambda (E)h (\Delta E,E), 
\end{equation} 
so that $\int_0^E h(\Delta E,E)d(\Delta E)=1$.

\subsection{Steady-state waiting-time distribution}\label{wtd}

\citet{dal07} showed how --- assuming a steady state --- the
waiting-time distribution for the jump transitions may be derived.
In this section we briefly re-iterate the theory, as it applies to
the present model.

Consider a deterministic trajectory described by
equation~(\ref{eq:stoch_de}), starting at energy $E_s$ and ending at
a higher energy $E_e$, the instant before a jump occurs. The distribution 
$p_e(E)$ of final
energies $E_e$ is given by the rate of jumping at a given energy 
divided by the mean total rate of jumping, i.e.\ 
\begin{equation}\label{eq:pee}
p_e(E)=\frac{\lambda (E)P(E)}{\langle \lambda\rangle }, 
\end{equation}
where $P(E)$ is the steady-state solution to the master
equation~(\ref{eq:master}) and 
\begin{equation}\label{eq:lambda_bar}
\langle\lambda \rangle =\int_0^{\infty}\lambda(E)P(E)dE 
\end{equation}
is the mean total rate. The distribution $p_s(E)$ of starting energies 
$E_s$ is then given by $p_e (E)$ together with the distribution of jumps
$h(\Delta E,E)$: 
\begin{equation}\label{eq:pse1}
p_s(E)=\int_{E}^{\infty}p_e(E^{\prime})h(E^{\prime}-E,E^{\prime})
dE^{\prime}, 
\end{equation} 
which using equations~(\ref{eq:h}) and~(\ref{eq:pee}) may be rewritten 
as 
\begin{equation}\label{eq:pse2}
p_s(E)=\frac{1}{\langle\lambda\rangle}
\int_{E}^{\infty}P(E^{\prime})\alpha (E^{\prime},E) dE^{\prime}.
\end{equation} 
The waiting-time distribution is given by
\begin{equation}\label{eq:wtd} 
p_{\tau}(\tau )=-\frac{d{\cal F}}{d\tau}
\end{equation} 
with 
\begin{equation}\label{eq:ftau} 
{\cal F}(\tau)=\int_0^{\infty}p_{\lambda}(E,\tau)dE, 
\end{equation} 
where $p_{\lambda}(E,t)$ is the solution to
\begin{equation}\label{eq:pde_plambda} 
\frac{\partial p_{\lambda}(E,t)}{\partial t}= 
-\frac{\partial }{\partial E}\left[\beta
(E)p_{\lambda}(E,t)\right] 
-\lambda (E)p_{\lambda}(E,t) 
\end{equation} 
with the initial condition $p_{\lambda}(E,0)=p_s (E)$.
Equation~(\ref{eq:pde_plambda}) describes the evolution of the system
before a flaring jump occurs, i.e.\ over the deterministic trajectory
starting at energy $E_s$ and ending at energy $E_e$.

A simpler form for the waiting-time distribution may be obtained when
$\beta (E)=\beta_0$, a constant. Solution of
equation~(\ref{eq:pde_plambda}) by characteristics then gives
\begin{equation} 
p_{\lambda }(E,t)=p_s(E-\beta_0 t)\exp \left\{
-\int_0^t\lambda \left[ E-\beta_0(t-s)\right] ds\right\}, 
\end{equation} 
assuming $E\geq \beta_0t$, and $p_{\lambda}(E,t)=0$ otherwise. In this
case 
\begin{equation} 
{\cal F}(\tau )=\int_0^{\infty}p_s (u)f(u,\tau)du
\end{equation} 
where 
\begin{equation}\label{eq:futau} f(u,\tau )=
\exp\left[-\int_0^{\tau}\lambda(\beta_0s+u)ds\right], 
\end{equation} 
so the waiting-time distribution is 
\begin{equation}\label{eq:ptau_beta0}
p_{\tau}(\tau )=\int_0^{\infty}p_s (u)\lambda (\beta_0\tau+u )f(u,\tau)
du. 
\end{equation}

\subsection{Steady-state analytic solutions}\label{analytic_solns}

Two analytic examples illustrate the application of the theory. The
examples are not relevant for flares, because they do not produce
power-law frequency-energy distributions, but they show how Poisson and
non-Poisson waiting-time distributions may be obtained.

The case $\beta (E)=\beta_0$ and $\alpha (E,E^{\prime})=\alpha_0$ (where
$\alpha_0$ and $\beta_0$ are constants) was considered by \citet{whe98}.
In this case equation~(\ref{eq:lambda}) gives $\lambda (E)=\alpha_0 E$, 
so the total rate of
jumps is energy dependent and the waiting-time distribution will not
correspond to a simple Poisson process. The analytic solution to the
steady-state master equation is 
\begin{equation}
P(E)=aEe^{-\frac{1}{2}aE^2}, 
\end{equation} 
with $a=\alpha_0/\beta_0$,
and from equation~(\ref{eq:ffe}) the frequency-energy distribution for
jumps is a Gaussian: 
\begin{equation} 
{\cal N}(E)=\alpha_0e^{-\frac{1}{2}aE^2}. 
\end{equation} 
From equation~(\ref{eq:lambda_bar}) the mean total rate is 
$\langle \lambda \rangle =\left(\alpha_0\beta_0\right)^{1/2}$, and 
using equations~(\ref{eq:pee}) and~(\ref{eq:pse2}) the distributions of 
end- and start-energies for deterministic trajectories are 
\begin{equation}
p_e(E)=\left(\frac{2}{\pi}\right)^{1/2}a^{3/2}E^2e^{-\frac{1}{2}aE^2}
\end{equation} 
and 
\begin{equation}\label{eq:pee_gauss}
p_s(E)=\left(\frac{2}{\pi}\right)^{1/2}a^{1/2}e^{-\frac{1}{2}aE^2}
\end{equation} 
respectively. Using equations~(\ref{eq:ptau_beta0})
and~(\ref{eq:pee_gauss}) it follows that the waiting-time distribution
is also a Gaussian: 
\begin{equation} 
p_{\tau}(\tau)=\left(\frac{2\alpha_0\beta_0}{\pi}\right)^{1/2}
e^{-\frac{1}{2}\alpha_0\beta_0\tau^2}. 
\end{equation}

As a second example, we consider the case $\beta (E)=\beta_0$ and
$\lambda (E)=\lambda_0$ (where $\beta_0$ and $\lambda_0$ are constants).
Since the total rate is constant the waiting-time distribution must be
\begin{equation}
\label{eq:wtd_poisson} 
p_{\tau}(\tau)=\lambda_0e^{-\lambda_0\tau}, 
\end{equation} 
i.e.\ jumps occur in time as a simple Poisson process. From 
equation~(\ref{eq:lambda}) this case requires 
$\alpha (E,E^{\prime})=\lambda_0/E$. The corresponding solution
to the steady-state master equation is 
\begin{equation}
P(E)=b^2Ee^{-bE}, 
\end{equation} 
with $b=\lambda_0/\beta_0$, and from
equation~(\ref{eq:ffe}) the frequency-energy distribution of jumps is
exponential: 
\begin{equation} 
{\cal N}(E)=b\lambda_0e^{-bE}. 
\end{equation}
The mean total rate of jumps given by equation~(\ref{eq:lambda_bar}) is
$\langle \lambda \rangle=\lambda_0$, and using equations~(\ref{eq:pee})
and~(\ref{eq:pse2}) we have $p_e(E)=b^2Ee^{-bE}$ and $p_s(E)=be^{-bE}$.
From equation~(\ref{eq:ptau_beta0}) it follows that the waiting-time
distribution is indeed given by equation~(\ref{eq:wtd_poisson}).

\subsection{Moments of the master equation}\label{moments}

Moments of the master equation give useful information about the global
behavior of solutions \citep{whelit01}.

The zeroth moment, obtained by integrating equation~(\ref{eq:master})
over all energies, gives the trivial result
\begin{equation}\label{eq:zeroth} 
\frac{d}{d t}\int_0^{\infty}P(E,t)dE=0, 
\end{equation} 
i.e.\ normalization is
preserved, provided $\beta (E,t)P(E,t)$ goes to zero as $E\rightarrow 0$
and $E\rightarrow \infty$.

The first moment, obtained by multiplying equation~(\ref{eq:master}) by
$E$ and integrating over all energies, gives the simple statement of
global energy balance: 
\begin{equation}\label{eq:first} 
\frac{d}{d t}\langle E\rangle =\langle \beta \rangle -\langle r\rangle ,
\end{equation} 
where for any quantity $f=f(E,t)$, the mean $\langle f \rangle$ is
defined by 
\begin{equation} 
\langle f\rangle = \int_0^{\infty}f(E,t)P(E,t)dE, 
\end{equation} 
and where
\begin{equation}\label{eq:r(E)} 
r(E) = \int_0^E (E-E^{\prime})\alpha
(E,E^{\prime}) dE^{\prime} 
\end{equation} 
is the total rate of energy loss at energy $E$. Equation~(\ref{eq:first}) 
requires $\beta (E,t)P(E,t)$ to go to zero as $E\rightarrow 0$ and 
$E\rightarrow \infty$. In the steady state equation~(\ref{eq:first}) gives
\begin{equation}\label{eq:first_steady} 
\langle \beta \rangle = \langle r \rangle . 
\end{equation}

\section{Flare-like solutions}\label{flare_solns}

In the following we consider solutions to the steady-state master
equation which may be of relevance to solar flares.

\subsection{Choices appropriate for flares}\label{flare_choices}

We restrict attention to the case $\beta (E)=\beta_0$, a constant. The
motivation is that active regions are externally driven, i.e.\ the
energy is supplied from the sub-photosphere by external processes. In the
absence of a back reaction, it is then expected that the energy supply
rate does not depend on the state of the system. In passing we note that
in general the energy supply rate may depend on time. However, in this
section we consider only steady-state solutions to the master equation.
We return to this point in \S\,4.

We consider the form 
\begin{equation}\label{eq:pl_alpha} 
\alpha (E,E^{\prime})=\alpha_0E^{\delta} (E-E^{\prime})^{-\gamma} \theta
(E-E^{\prime}-E_c) 
\end{equation} 
for the flare transition rate, where
$E_c$ is a low-energy cutoff, and $\theta (x)$ is the step function. The
case $\delta=0$ was considered in \cite{whe98}. The motivation for
equation~(\ref{eq:pl_alpha}) is that it may describe an avalanche-type
system, in which energy transitions are intrinsically power-law
distributed. The power law is assumed to originate in the microphysics
of the flare process, and must be assumed at the level of this model.
[This is in contrast to models which attempt
to account for the power law, e.g.\ \cite{ros78}.]
The low-energy cutoff $E_c$ is needed to ensure $\lambda (E)$ is finite.
The $E^{\delta}$ factor represents a possible dependence of the
transition rate on the energy of the system. It is plausible that an
avalanche-type system with more energy is more likely to contain
unstable sites, and hence will flare at a higher rate. In the following
we take $\gamma=1.5$ in every instance, and consider two cases:
$\delta=0$, following \citet{whe98}; and $\delta=1$.
The choice of transition rates~(\ref{eq:pl_alpha}) leads to the total
flaring rate
\begin{equation}\label{eq:lambda_pl_alpha} 
\lambda (E)=\frac{\alpha_0}{\gamma-1}E^{\delta}\left( E_c^{-\gamma+1}
-E^{-\gamma+1}\right). 
\end{equation} 
Hence the rate of occurrence of
flares is energy-dependent, and the waiting-time distribution will not
correspond to a simple Poisson process, as pointed out in \S\,1.

Application of the first moment with the choice of transition 
rates~(\ref{eq:pl_alpha}) leads to a simple estimate
for the mean energy of the system \citep{whe02}. Specifically, from
equation~(\ref{eq:r(E)}) we have 
\begin{equation} 
r(E)\approx
\frac{\alpha_0}{2-\gamma}E^{\delta+2-\gamma}, 
\end{equation} 
for $E\gg E_c$. Taking averages and making the approximation 
$\langle E^{\delta+2-\gamma}\rangle \approx \langle 
  E\rangle^{\delta+2-\gamma}$
together with $\langle\beta \rangle=\beta_0$ in 
equation~(\ref{eq:first_steady}) gives
\begin{equation}\label{eq:ebar_pl_alpha} 
\langle E\rangle \approx
\left(\frac{2-\gamma}{\alpha_0/\beta_0}
\right)^{\frac{1}{\delta+2-\gamma}}. 
\end{equation}

Substituting equation~(\ref{eq:pl_alpha}) into equation~(\ref{eq:ffe})
leads to the form for the flare frequency-energy distribution
\begin{equation}\label{eq:ffe_pl_alpha} 
{\cal N}(E)=\alpha_0 E^{-\gamma}
\int_{E}^{\infty}\left(E^{\prime}\right)^{\delta}
P(E^{\prime})dE^{\prime} 
\end{equation} 
for $E\geq E_c$. Hence it
follows that the frequency-energy distribution will be a power law with
index $\gamma$ up to energies $E$ at which $P(E)$ becomes very
small.  Equation~(\ref{eq:ebar_pl_alpha}) provides a crude estimate
(a lower bound) for the energy at which the frequency-energy distribution 
is expected to depart from power law behavior.

\subsection{Numerical method}\label{numerical}

The steady-state master equation may be non-dimensionalized by
introducing new variables ${\overline E}=E/E_0$, ${\overline P}=PE_0$,
${\overline t}=\beta_0 t/E_0$, and ${\overline \alpha }= E_0^2 \alpha
/\beta_0$, where $E_0$ is a chosen scale for energy. [For the solutions
corresponding to equation~(\ref{eq:pl_alpha}) we take $E_0=E_c$.]
This procedure gives 
\begin{equation}\label{eq:steady_master_ndim}
\frac{d\overline P}{d\overline E} +{\overline \lambda }{\overline P}
-\int_{\overline E}^{\infty} {\overline P}({\overline E}^{\prime})
{\overline \alpha} ({\overline E}^{\prime},{\overline E})
d{\overline E}^{\prime}=0, \end{equation} where \begin{equation}
{\overline \lambda }= \int_0^{\overline E} {\overline \alpha }
({\overline E},{\overline E}^{\prime}) d{\overline E}^{\prime}.
\end{equation} 
Hereafter we assume non-dimensional equations, but omit the bars.

Equation~(\ref{eq:steady_master_ndim}) is linear in $P(E)$, and hence
may be solved by discretizing in energy and solving a coupled system of
linear equations. These equations must be supplemented by the
normalization condition on $P(E)$. Direct back-substitution provides an
efficient method of solution, and the details of the procedure are given
in Appendix A. In \citet{whe98} the steady-state master equation was
solved by a relaxation procedure, but the approach given here is more
numerically efficient.

The flare frequency-energy distribution is obtained from the solution
for $P(E)$ via numerical evaluation of equation~(\ref{eq:ffe}). The
waiting-time distribution is similarly determined via numerical
evaluation of equation~(\ref{eq:ptau_beta0}), using an analytic form
for $f(u,\tau)$ obtained from equation~(\ref{eq:futau}). All numerical
integrations use the extended trapezoidal rule. The numerical solution
was tested on the analytic cases given in \S\,\ref{analytic_solns}.

\subsection{Results}\label{results}

\begin{figure} \epsscale{.80} \plotone{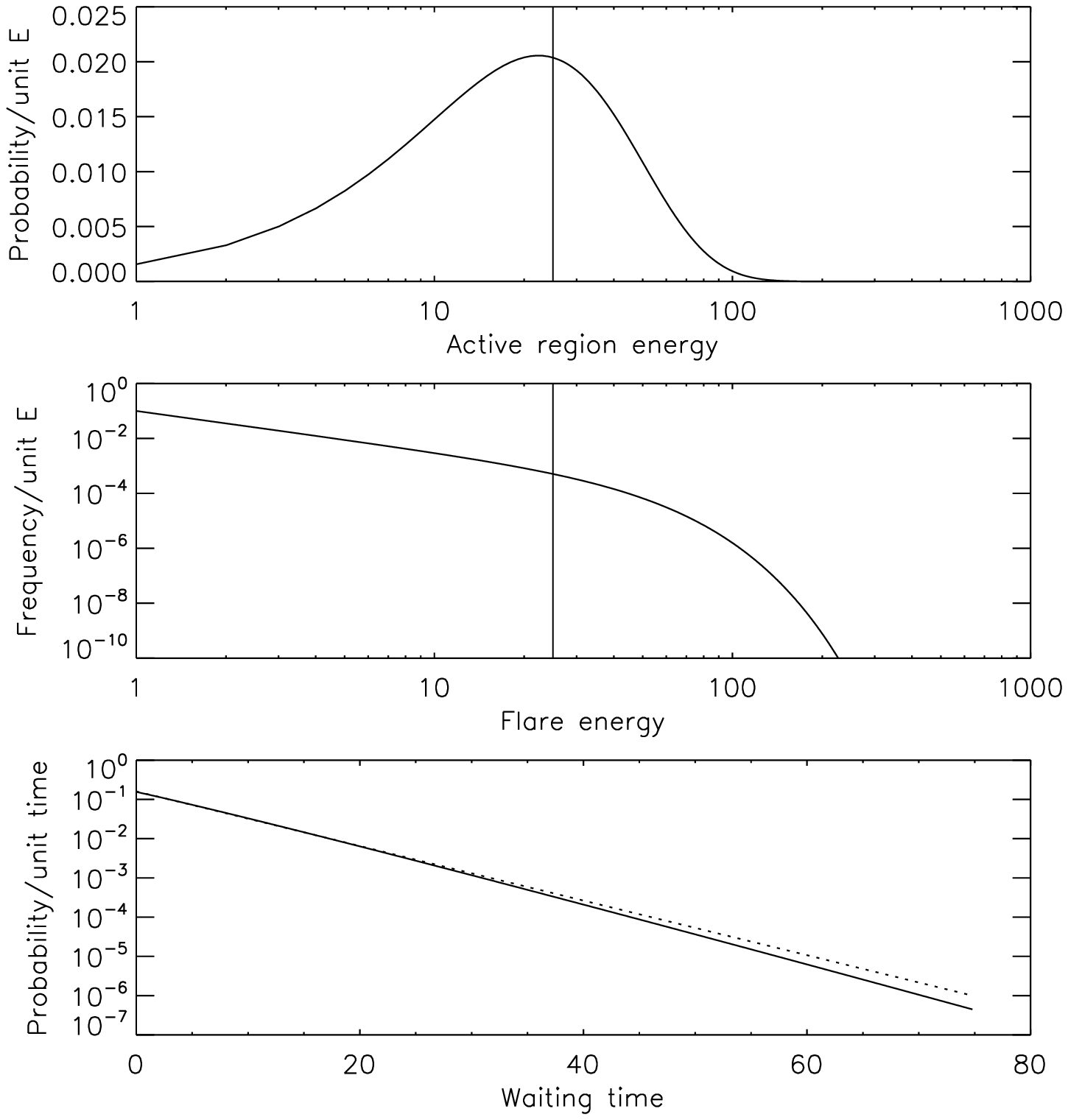} \caption{Numerical
solution to the steady-state master equation for the case $\delta=0$,
$\gamma=1.5$, and $\alpha_0=0.1$, one of the cases considered in 
\citet{whe98}. Upper panel: probability distribution
for free energy $P(E)$; middle panel: flare frequency-energy 
distribution ${\cal N}(E)$;
lower panel: flare waiting-time distribution $p_{\tau}(\tau)$.\label{fig1}} 
\end{figure}

First we consider the case $\delta = 0$, following \citet{whe98}.
Figure~\ref{fig1} illustrates the numerical solution of the steady-state
master equation~(\ref{eq:steady_master_ndim}) for the case $\alpha_0
=0.1$, which is one of the two cases considered in \citet{whe98}. The
upper panel shows the probability distribution $P(E)$ for active region
energy (as a linear-log plot), the middle panel shows the flare
frequency-energy distribution ${\cal N}(E)$ (as a log-log plot), and the
lower panel shows the waiting-time distribution $p_{\tau}(\tau )$ (as a
log-linear plot). As explained in \S\,\ref{flare_choices}, the
frequency-energy distribution is expected to be a power law with index
$\gamma=1.5$ below energies at which $P(E)$ becomes small, and the
expression~(\ref{eq:ebar_pl_alpha}) provides a lower bound for the 
departure from power-law behavior. The lower bound is shown in the
upper and middle panels by a vertical line. The upper and middle panels
confirm the results of \citet{whe98}. The lower panel shows the
waiting-time distribution $p_{\tau} (\tau )$ (solid curve) as well as
the Poisson distribution $\langle \lambda\rangle e^{-\langle
\lambda\rangle\tau}$ (dotted line) corresponding to the mean rate of
flaring implied by the form of $\lambda (E)$ and the solution for $P(E)$
[see equation~(\ref{eq:lambda_bar})]. Note that the units for time in
the lower panel are $E_c/\beta_0$, following the non-dimensionalization
in \S\,\ref{numerical}. The waiting-time distribution for the model is
approximately Poisson, although there is a slight deficiency of long
waiting-times by comparison with the Poisson distribution.

\begin{figure} \epsscale{.80} \plotone{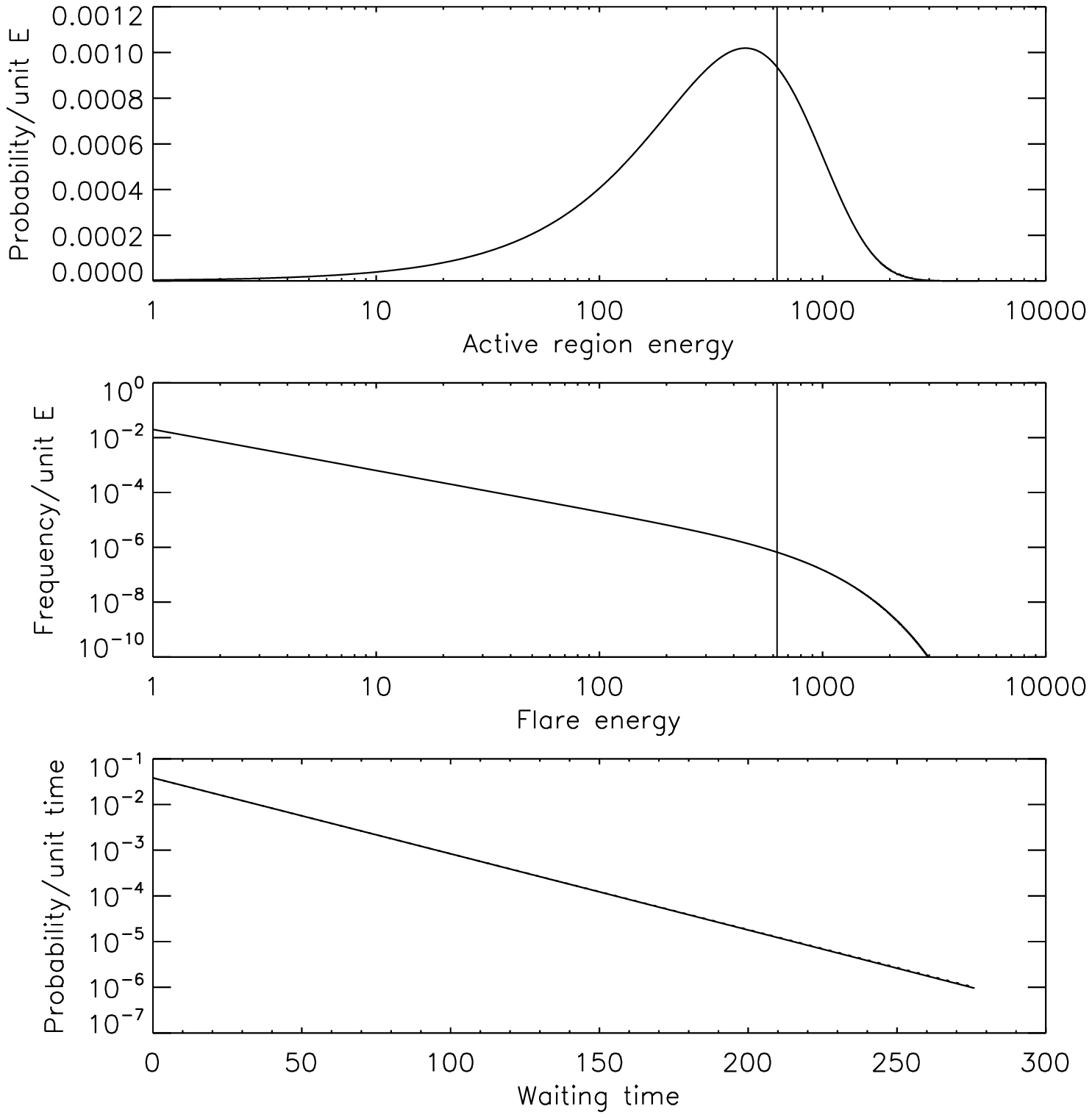} \caption{Numerical
solution to the steady-state master equation for the case $\delta=0$,
$\gamma=1.5$, and $\alpha_0=0.02$, one of the cases considered in 
\citet{whe98}. Upper panel: probability distribution
for free energy $P(E)$; middle panel: flare frequency-energy 
distribution ${\cal N}(E)$;
lower panel: flare waiting-time distribution $p_{\tau}(\tau)$.\label{fig2}} 
\end{figure}

Figure~\ref{fig2} shows the case $\delta=0$ and $\alpha_0=0.02$, which is
the other case considered by \citet{whe98}, in the same format as
Figure~\ref{fig1}. For a lower value of $\alpha$ the system flares less
often and hence is more likely to have larger energy. Hence the distribution
$P(E)$ shown in the upper panel is shifted to higher energy and has a
higher mean. The flare frequency-energy distribution (middle panel) is a
power law over more decades in energy than for the case $\alpha_0=0.1$. 
These results are consistent with the findings of \citet{whe98}.
The lower panel shows the waiting-time distribution (solid curve) as 
well as the Poisson distribution corresponding to 
$\langle \lambda \rangle$ (dotted), although the two curves are almost
indistinguishable. 

\begin{figure} \epsscale{.80} \plotone{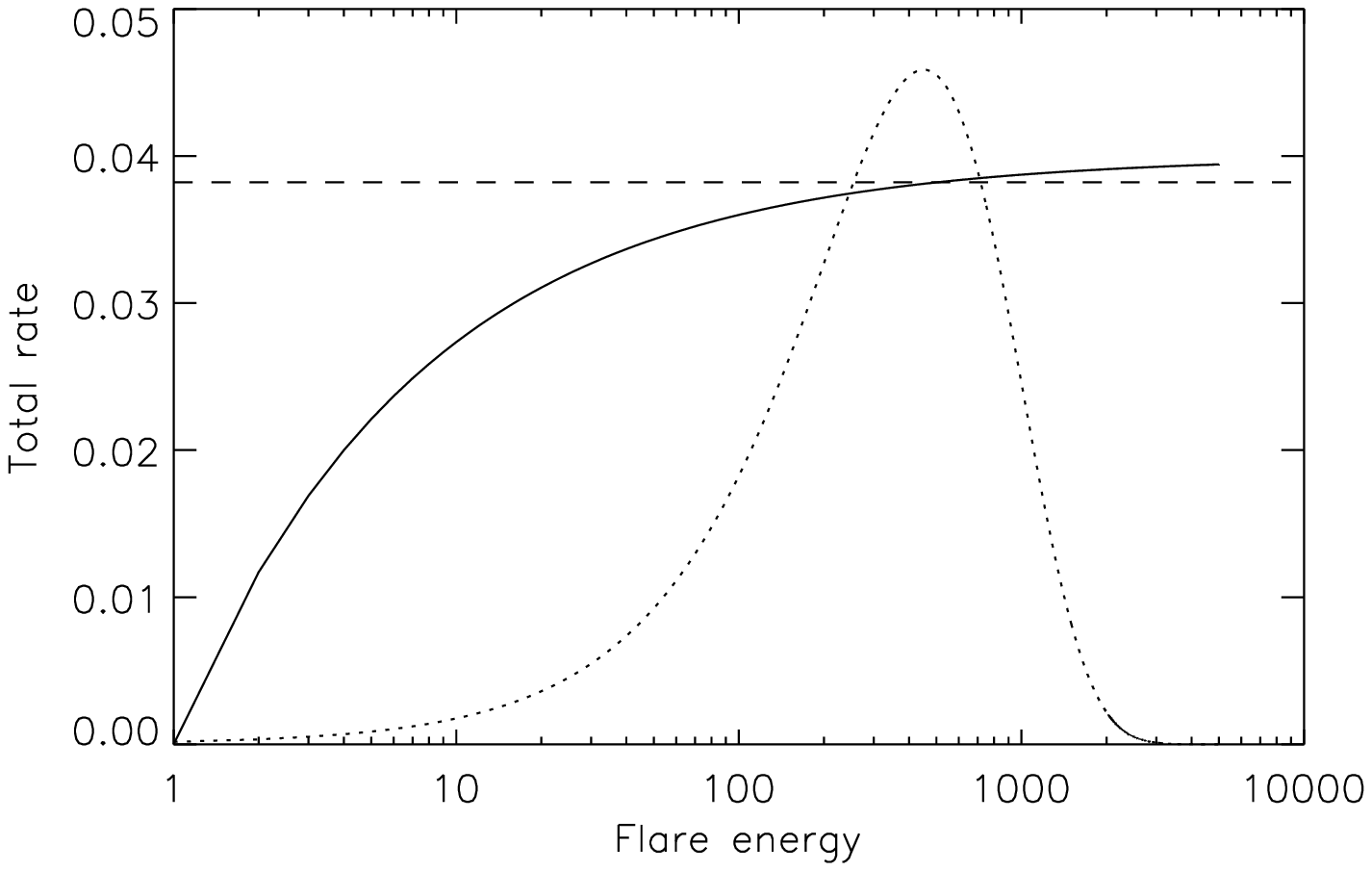} \caption{The total rate
of flaring $\lambda (E)$ versus energy (solid curve) for the case 
$\delta=0$, $\gamma=1.5$, and $\alpha_0=0.02$, and the mean
total rate $\langle \lambda \rangle$ (dashed line). The energy
distribution $P(E)$ is also shown, with an arbitrary normalization
(dotted curve).\label{fig3}} 
\end{figure}

The results in Figures~\ref{fig1} and~\ref{fig2} suggest that the
$\delta=0$ model has a waiting-time distribution which is close to being
strictly Poisson (exponential), and that the approximation becomes
better for smaller values of $\alpha_0$. This may be understood in terms
of the expression~(\ref{eq:lambda_pl_alpha}). For $E\gg E_c$ the total
rate is $\lambda (E)\approx \alpha_0 E_c^{-\gamma+1}/(\gamma-1)$, which
is constant, in which case a Poisson waiting-time distribution is 
expected. For smaller values of $\alpha_0$ the system is more likely
to have larger energy, and hence the approximation $E\gg E_c$ will be
better. Figure~\ref{fig3} illustrates this explanation for the case
$\delta=0$, $\alpha_0=0.02$. The solid curve shows the total rate as a
function of energy, and the dashed line shows the mean total rate
$\langle\lambda\rangle$. The energy distribution $P(E)$ is shown, with
arbitrary normalization, by the dotted curve. We see that, over the
range of energy for which the distribution $P(E)$ is substantial, the
total rate is approximately constant and equal to the mean total rate. 
These results resolve the puzzle identified in
\S\,\ref{intro}: the waiting-time distribution for the model is not
strictly Poisson, but is a good approximation to an exponential.

\begin{figure} \epsscale{.80} \plotone{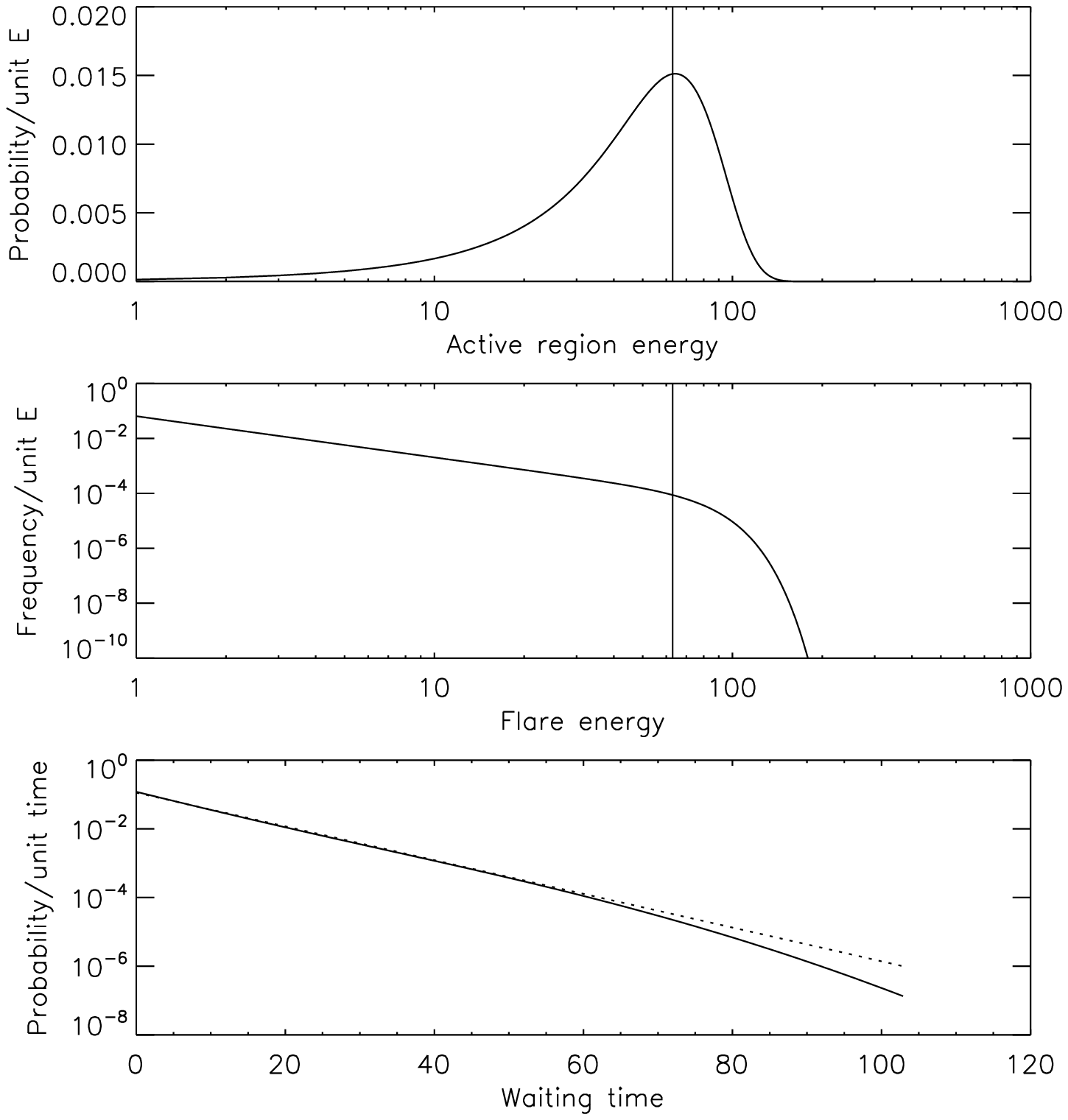} \caption{Numerical
solution to the steady-state master equation for the case $\delta=1$,
$\gamma=1.5$, and $\alpha_0=10^{-3}$. Upper panel: probability distribution
for free energy $P(E)$; middle panel: flare frequency-energy distribution
${\cal N}(E)$;
lower panel: flare waiting-time distribution $p_{\tau}(\tau)$.\label{fig4}} 
\end{figure}

\begin{figure} \epsscale{.80} \plotone{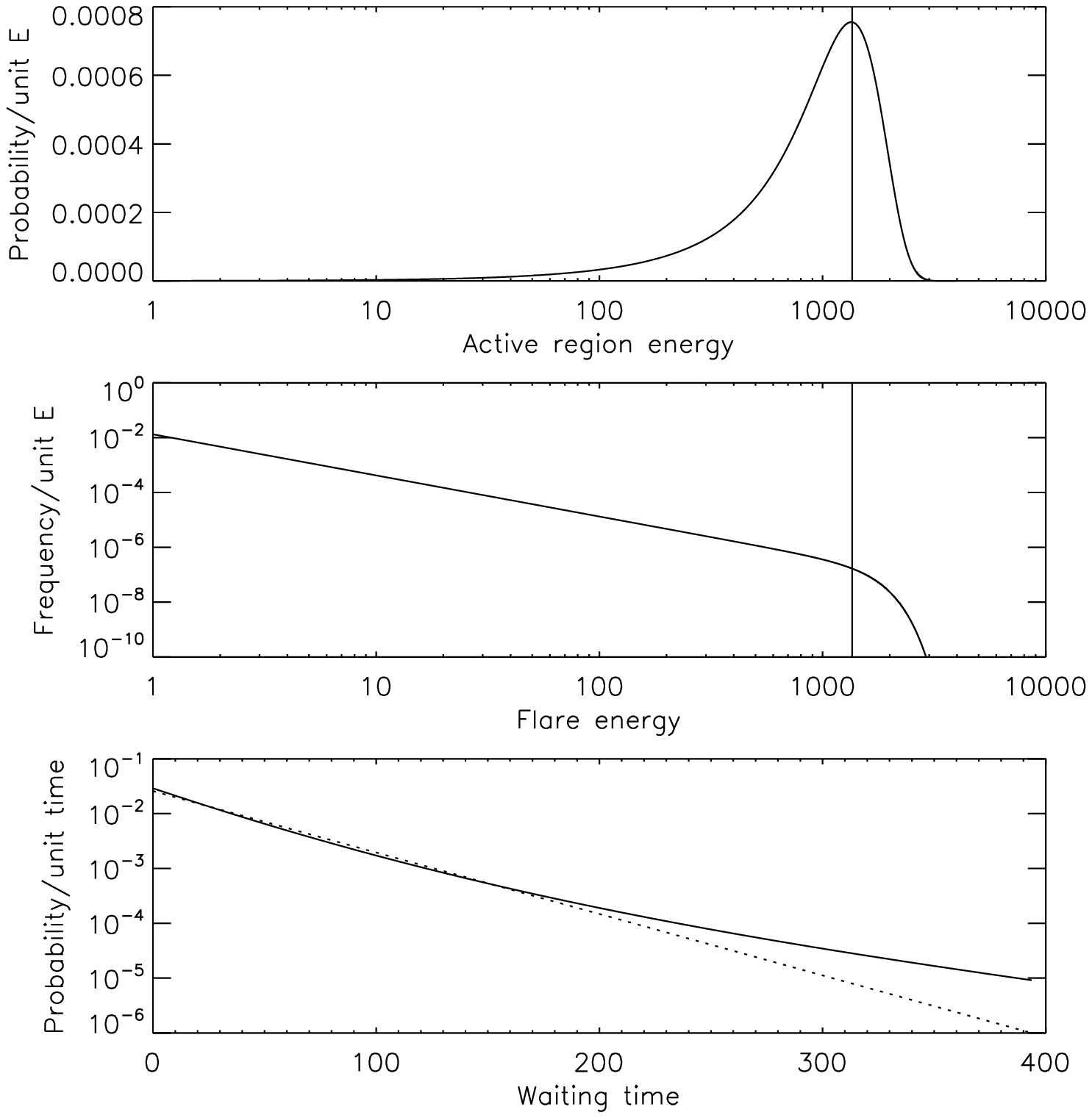} \caption{Numerical
solution to the steady-state master equation for the case $\delta=1$,
$\gamma=1.5$, and $\alpha_0=10^{-5}$. Upper panel: probability distribution
for free energy $P(E)$; middle panel: flare frequency-energy distribution
${\cal N}(E)$;
lower panel: flare waiting-time distribution $p_{\tau}(\tau)$.\label{fig5}} 
\end{figure}

Next we consider the case $\delta=1$, to examine what happens when the
rate of flare transitions increases with the energy of the system.
Figures~\ref{fig4} and~\ref{fig5} show the cases $\alpha_0=10^{-3}$ and
$\alpha_0=10^{-5}$ respectively, in the same format as Figures~1 and~2.
First consider Figure~\ref{fig4}. The energy distribution $P(E)$ shown
in the upper panel is qualitatively similar to the $\delta=0$ 
case, although the distribution declines more rapidly at higher energies, 
so that it is more skewed in a linear-log representation. Since the rate 
of transitions increases with energy, the system is less likely to be found 
at very large energies, and this explains the rapid decline. The middle 
panel shows the flare frequency-energy distribution ${\cal N}(E)$,
which is a power law with index $\gamma$ below a high energy rollover.
The estimate~(\ref{eq:ebar_pl_alpha}) for the mean of the distribution
(vertical line) provides a lower bound for the departure from power law 
behavior. The lower panel shows the waiting-time distribution 
$p_{\tau} (\tau )$, and the Poisson distribution implied by
$\langle\lambda\rangle$. The waiting-time distribution is approximately
Poisson, but has a deficit of long waiting times. Figure~\ref{fig5}
illustrates the case with reduced flare transition rates. 
The distribution
$P(E)$ (upper panel) is shifted to higher energies, and is again quite
skewed in the linear-log representation. The frequency-energy
distribution ${\cal N}(E)$ (middle panel) is a power law over more
decades in energy. The waiting-time distribution $p_{\tau}(\tau )$
(lower panel) is again approximately exponential, but departs somewhat
from the Poisson model, including showing an excess of long waiting
times.

\begin{figure} \epsscale{.80} \plotone{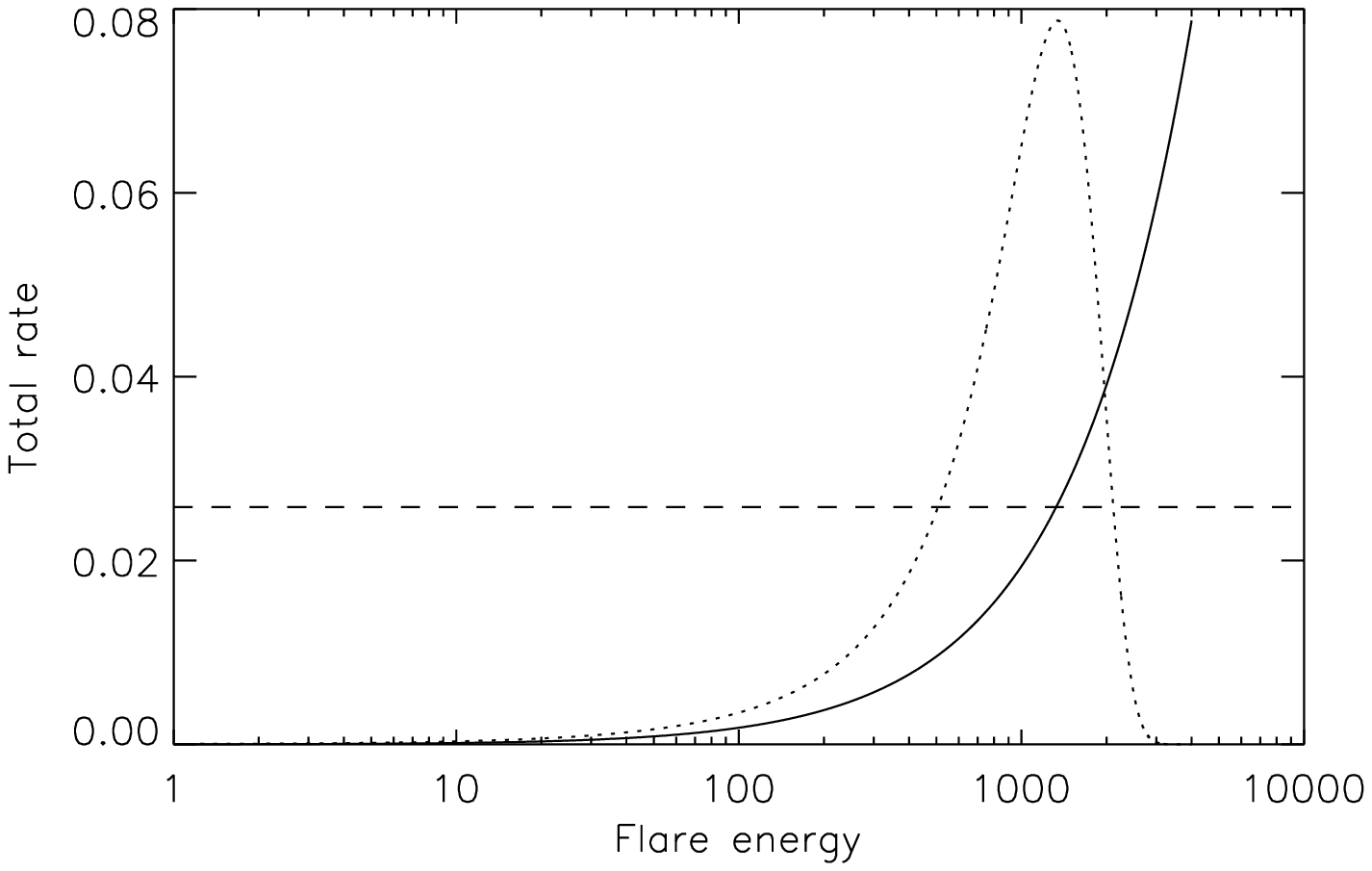} \caption{The total rate
of flaring $\lambda (E)$ versus energy (solid curve) for the case 
$\delta=1$, $\gamma=1.5$,
and $\alpha_0=0.02$, and the mean
total rate $\langle \lambda \rangle$ (dashed line). The energy
distribution $P(E)$ is also shown, with an arbitrary normalization
(dotted curve).\label{fig6}} \end{figure}

The results in the lower panels of Figures~\ref{fig4} and~\ref{fig5}
suggest that, for the $\delta=1$ model, the waiting-time distribution is
approximately exponential but shows some departure from the Poisson case
depending on the parameters of the model. The approximate Poisson
behavior is perhaps surprising because in this case the total rate of
flaring [given by equation~(\ref{eq:lambda_pl_alpha})] varies
approximately linearly with $E$ (for $E\gg E_c$). Figure~\ref{fig6}
illustrates this for the case $\delta=1$, $\alpha_0=10^{-5}$, using the
same format as Figure~\ref{fig3}. The total rate (solid curve) increases
substantially over the range of energies the system is likely to have
[the dotted curve shows $P(E)$], and may be substantially different to
$\langle\lambda\rangle$ (the dashed line). However, $p_{\tau} (\tau )$
is defined by a complicated average of the rate over energy, which is
different for different waiting times [see
equation~(\ref{eq:ptau_beta0})], and the numerical results show that the
resulting waiting-time distribution is approximately Poisson.

\section{Discussion}\label{discussion}

A stochastic model is presented for the free magnetic energy of a flaring 
solar active region. The energy of an active region is assumed to grow 
deterministically between random flare events at which the energy jumps 
downwards by an amount equal to the flare energy. Flares jumps occur from 
energy $E$ to $E^{\prime}$ with a rate $\alpha (E,E^{\prime})$ per unit 
time and per unit energy, and energy input occurs at a rate $\beta (E)$. 
Active region energy is then described by a distribution $P(E)$ which 
is the steady state solution to a master equation. This distribution 
determines two observable distributions, namely the flare frequency-energy 
distribution, and the waiting-time distribution. The model generalizes 
and extends the approach of \citet{whe98}. Novel aspects of the work 
presented here include the 
determination of waiting-time distributions [following general theory 
presented by \citet{dal07}], consideration of a more general form for the 
rate of flare transitions, and introduction of an efficient method of 
numerical solution of the steady-state master equation.

The form $\alpha (E,E^{\prime})
=\alpha_0E^{\delta}(E-E^{\prime})^{-1.5}$ for flare transitions is
investigated, for the cases $\delta=0$ and $\delta =1$. The case $\delta
= 0$ was considered by \citet{whe98}, motivated by the avalanche model.
For both cases the model is shown to produce power-law flare
frequency-energy distributions below a rollover at high energies, due to
the active region having a finite energy. For both cases the
waiting-time distribution is approximately exponential (Poisson). For
the case $\delta = 0$ this may be understood in that the total rate
$\lambda (E)$ is approximately constant for $E\gg E_c$, which becomes a
good approximation for small $\alpha_0$, when the system is more likely
to be found at large energies. This result is consistent with the
interpretation of this model as avalanche-like, since avalanche models
have simple Poisson statistics. For the case $\delta =1$ the
interpretation is more complicated, because the total rate varies
approximately linearly with energy. However, the waiting-time
distribution is determined by an average of rates over the possible
energies of the active region, and numerical evaluation shows that the
result is approximately exponential.

The general model introduced in \S\,\ref{master} includes time
dependence in the driving rate, but we have focused on the steady state
throughout this paper. Many active regions exhibit large variations in the 
rate of flaring during a transit of the disk \citep{whe01}. This 
behavior is often linked e.g.\ with the emergence of new magnetic flux
\citep{rom07}, which suggest that it is a response to an increased rate 
of driving.  Hence we have neglected an important aspect of active region 
energetics.  Time-dependent driving will influence the observed 
waiting-time distribution.
In the simplest case time variation might be represented by a piecewise
constant variation in the driving rate. If the system adjusts suitably
quickly to changes in driving, the steady state solution applies to each
piece. The waiting-time distribution is then a weighted sum over the
steady-state distributions applying to each piece \citep{whelit01}. 
Based on the
results presented in this paper, the waiting-time distribution for
active regions is expected to be approximately exponential provided the
rate of driving of the system is constant. If the rate of driving is
time varying, the distribution will depart from exponential. A
time-dependent model will be investigated in future work.

One shortcoming of the model is that it does not describe energy loss
from the system by mechanisms other than flaring, for example loss due
to flux submergence, or quasi-steady background dissipation. However, a
simple generalization of the master equation permits this. Specifically,
if the secular energy `input' is replaced by small jumps in energy
(which may be positive or negative), then the energy gains and losses
may be represented by Fokker-Planck terms in a generalized master or
Chapman-Kolmogorov equation. Specifically, the secular energy increase 
term $-\partial \left[\beta (E,t)P(E,t)\right]/\partial E$
on the right hand side of equation~(\ref{eq:master}) may be replaced
by a pair of terms 
$-\partial \left[\beta_1 (E,t) P(E,t)\right]/\partial E
+\frac{1}{2}\partial^2\left[\beta_2 (E,t) P(E,t)\right]/\partial E^2$,
where the coefficients $\beta_i (E,t)$, $i=1,2$ represent first and second
moments of energy changes associated with the small jump transitions
\citep{vankam92,gar04}. It is straightforward to solve the resulting
generalize master equation by discretization and solution of the 
resulting linear system, extending the approach presented in the 
Appendix. However, in this case the \citet{dal07} method for determining 
the steady-state waiting-time distribution needs to be modified. 
This model will be investigated more completely in future work.

The results presented here show how it is possible to construct
a model for active region energetics which directly predicts observable 
flare statistics, namely the flare frequency-energy, and waiting-time 
distributions. In principle the observations may be used to determine
the energy supply and flare energy release terms in the model, i.e.\ 
$\beta (E,t)$ and $\alpha (E,E^{\prime})$. However, the observations 
are not really precise or unambiguous enough to identify these terms 
with certainty. In particular, the interpretation of the waiting-time 
distribution is complicated by the time dependence of the energy 
supply. The situation would be improved by an ability to estimate coronal 
free energy and the rate of supply of energy to active regions from 
observations. Reliable methods for estimating these quantities are the
subject of current research \citep{wel07,sch08}. If such methods are
developed, the theory developed here will be of greater significance. It
may provide valuable insight into the flare mechanism, as well as being
of practical benefit for flare prediction.

\acknowledgments

The author thanks Ian Craig for pointing out that the master equation
can be solved as a linear system by back-substitution, and a referee
for helpful comments which improved the presentation.

\appendix

\section{Appendix} \section*{Numerical solution of the steady-state
master equation}

Discretizing equation~(\ref{eq:steady_master_ndim}) at energies
$E_i=i\Delta$ gives 
\begin{equation}\label{eq:master_discrete}
\frac{P_{i+1}-P_{i-1}}{2\Delta}
+\lambda_iP_i-\frac{1}{2}\Delta\sum_{j=i}^{N-2}\left(P_j\alpha_{j,i}
+P_{j+1}\alpha_{j+1,i}\right) =0 
\end{equation} 
where 
\begin{equation}
\lambda_i = \frac{1}{2}\Delta \sum_{j=0}^{i-1}\left(
\alpha_{i,j}+\alpha_{i,j+1}\right) 
\end{equation} 
and where $P_i=P(E_i)$
and $\alpha_{i,j}=\alpha(E_i,E_j)$. Centered differencing is used for
the derivative, and the extended trapezoidal rule is used for the
integrals. Equation~(\ref{eq:master_discrete}) with $i=1,2,...,N-1$ (and
the assumption $P_{N}=0$) may be supplemented by the normalization
condition 
\begin{equation}\label{eq:norm_discrete}
\frac{1}{2}\Delta\sum_{i=0}^{N-2}\left(P_i+P_{i+1}\right) =1
\end{equation} 
to give $N$ linear equations in the $N$ unknowns
$P_0,P_1,P_2,...,P_{N-1}$.

The resulting linear system may be solved efficiently by
back-substitution as follows. Equation~(\ref{eq:master_discrete}) may be
re-written 
\begin{equation}\label{eq:master_discrete2}
P_{i-1}=P_{i+1}+2\Delta \left[ \lambda_i P_i
-\frac{1}{2}\Delta\sum_{j=i}^{N-2}\left(P_j\alpha_{j,i}
+P_{j+1}\alpha_{j+1,i}\right)\right], 
\end{equation} 
which expresses
$P_{i-1}$ in terms of $P_{i},P_{i+1},P_{i+2},...,P_{N-2}$. Hence if we
assume a value for $P_{N-1}$, we can apply
equation~(\ref{eq:master_discrete2}) to solve for $P_{N-2}$, and then
apply it again to solve for $P_{N-3}$, etc. In this way we can determine
$P_{N-1},P_{N-2},P_{N-3},...,P_{0}$ up to an unknown normalization
factor. The factor is determined by applying
equation~(\ref{eq:norm_discrete}). Specifically, the solution is given
by $P_i^{\prime}$ (with $i=0,1,2,...,N-1$), where 
\begin{equation}
P_i^{\prime}=\frac{P_i}{
\frac{1}{2}\Delta\sum_{i=0}^{N-2}\left(P_i+P_{i+1}\right)}.
\end{equation}

\clearpage

\clearpage

\clearpage

\clearpage

\clearpage

\clearpage

\end{document}